\input{aipcheck}

\documentclass[
    ,final            
  ]
  {aipproc}

\layoutstyle{6x9}
\usepackage{epsfig}

\begin{document}

\title{The Acceleration of Electrons at Perpendicular Shocks and its Implication for Solar Energetic Particle events}


\author{Fan Guo}{
}

\author{Joe Giacalone}{
  address={Department of Planetary Sciences, University of Arizona, Tucson, AZ 85721}
}


\begin{abstract}
We present a study of the acceleration of electrons at a perpendicular shock that propagates through a turbulent
magnetic field. The energization process of electrons is investigated by utilizing a combination of hybrid (kinetic
ions and fluid electron) simulations and test-particle electron simulations. In this method, the motions of the test-particle electrons are numerically integrated in the time-dependent electric and magnetic fields generated by two-dimensional hybrid simulations. We show that large-scale magnetic fluctuations effect electrons in a number of ways and
lead to efficient and rapid energization at the shock front. Since the electrons mainly follow along magnetic lines of
force, the large-scale braiding of field lines in space allows the fast-moving electrons to interact with the shock
front and get accelerated multiple times. Ripples in the shock front occurring at various scales will also contribute
to the acceleration by mirroring the electrons. Our calculation shows that this process favors electron
acceleration at perpendicular shocks. The acceleration efficiency is critically dependent on the turbulence amplitude
and coherence length. We also discuss the implication of this study for solar energetic particles (SEPs) by comparing
the acceleration of electrons with that of protons. Their correlation indicates that perpendicular shocks play an important
role in SEP events.
\end{abstract}

\maketitle


\section{Introduction}
The acceleration of charged particles at collisionless shocks is a remarkable process in astrophysics and space physics. The diffusive shock acceleration (DSA) has been very successful in explaining the acceleration process and is thought to be the mechanism for accelerating solar energetic particles (SEPs). However, how a population of low energy particles gain energy, the so-called ``injection problem'', is still not well understood. The acceleration of electrons at collisionless shocks is considered to be more difficult than that of ions. This is primarily due to the fact that the gyroradii of electrons are much smaller compared with that of protons at the same energy (by a factor of $\sqrt{m_i/m_e} \sim 43$), therefore low energy electrons can not resonantly interact with large-scale turbulence or ion-scale waves. In contrast, energetic electrons are often observed to be associated with quasi-perpendicular shocks.

In SEP events, energetic electrons are found to be associated with energetic ions. \citet{Haggerty2009} reported that energetic electrons (175-315 keV) are well associated with protons with energy between $1.8$ and $4.7$ MeV. \citet{Cliver2009} showed that, the electrons ($\sim 0.5$ MeV) and the protons ($>10$ MeV) in large SEP events are strongly correlated. The correlation still exists even when the isotope ratio (Fe/O) changes. The tight correlation between electrons and ions has also been found in ground base level events (Tylka, private communication \citep{Tylka2012}), where both electrons and protons are accelerated to relativistic energies. It is interesting to note that the high energy electrons and protons are also closely correlated in solar flares \citep{Shih2009}. These observations indicate that a common mechanism may exist in the process of particle acceleration during SEP events.

Particle acceleration at perpendicular shocks has been considered as a fast and efficient process \citep{Jokipii1982,Jokipii1987}. \citet{Giacalone2005a,Giacalone2005b} has performed numerical simulations for the acceleration of  protons at fast shocks that propagates through a plasma that contains large scale magnetic fluctuations. The acceleration to high energy is very efficient, which indicates that there is no injection problem. Recently, \citet{Jokipii2007} have proposed a mechanism to solve the injection problem that does not require strong pitch-angle scattering from small-scale fluctuations. The fast-moving electrons can move along meandering magnetic field lines and travel back and forth across the shock front, and, therefore gain energy from the difference between the upstream and downstream flow velocities. Using self-consistent hybrid simulations combined with test-particle simulations for electrons, \citet{Guo2010} have found efficient electron acceleration at perpendicular shocks moving through a plasma containing large-scale pre-existing upstream magnetic turbulence. The turbulent magnetic field leads to field-line meandering that allows the electrons to get accelerated at the shock front multiple times, like in the Jokipii \& Giacalone picture. In a more recent paper, \citet{Guo2012} demonstrated that in solar flare region, a low Mach number shock can also efficiently accelerate both electrons and ions.
In this paper, we present the results of the acceleration of electrons at a perpendicular shock that propagates through a turbulent magnetic field. The energization process of electrons is investigated by utilizing a combination of hybrid simulations and test-particle electron simulations.  We also discuss the implication of this study for SEPs by comparing the acceleration of electrons with that of protons. Their correlation indicates that perpendicular shocks play an important role in SEP events.

\section{Numerical Method}

We implement a combination of a $2$-D
hybrid simulation to model the fields and plasma flow and a test
particle simulation to follow the orbits of electrons. In the first step, we employ a two-dimensional
hybrid simulation 
that includes pre-existing large-scale turbulence \citep{Giacalone2005b}. In the hybrid
simulation model, the ions are treated
fully kinetically and thermal (i.e., non-energetic) electrons are
treated as a massless fluid. This approach is well suited to resolve
ion-scale plasma physics that is critical to describe supercritical
collisionless shocks. We consider a two-dimensional
Cartesian grid in the $x-z$ plane. All the physical vector
quantities have components in three directions, but depend spatially
only on the two variables. A shock is produced by continuously injecting 
plasma from one end ($x=0$) of the simulation box, and
reflected elastically at the other end ($x=L_x$). This boundary is
also assumed to be a perfectly conducting barrier. The pileup of
density and magnetic field creates a shock propagating in the $-x$
direction. To include the large-scale magnetic fluctuations, a
random magnetic field is superposed on a mean field at the beginning
of the simulation and is also injected continuously at the $x=0$
boundary during the simulation. The simplified one-dimensional
fluctuations have the form $\textbf{B}(z, t) = \delta \textbf{B}(z,
t) + \textbf{B}_1$, where $\textbf{B}_1$ is the averaged upstream
magnetic field. The fluctuating component contains an equal mixture
of right- and left-hand circularly polarized, forward and backward
parallel-propagating plane Alfven waves. The amplitude of the
fluctuations is determined from a Kolmogorov-like power spectrum:
$$
P(k) \propto \frac{1}{1 + (k L_c)^{5/3}},
$$
\noindent in which $L_c$ is the coherence scale of the fluctuations. For the simulations presented in
this study, we take $L_c = L_z$, where $L_z$ is the size of simulation box in the $z$
direction. As a standard case of this study, 
we consider a turbulence variance $\sigma = \delta B^2/B_1^2 = 0.3$.
The size of the simulation box for most of situations is 
$L_x\times L_z = 400 c/\omega_{pi}\times 1024 c/\omega_{pi}$, 
where $c/\omega_{pi}$ is the ion inertial length. The Mach
number of the flow in the simulation frame is $M_{A0} = 4.0$, and the averaged Mach
number in the shock frame is about $5.6$. The averaged shock normal angle 
$\langle \theta_{Bn} \rangle = 90^\circ$, but we also present the cases for $\langle \theta_{Bn} \rangle = 60^\circ$ and $75^\circ$. We also discuss the effect of different values of turbulence variances and correlation lengths. The other important simulation parameters include electron and ion plasma beta $\beta_e =
0.5$ and $\beta_i = 0.5$, respectively, grid sizes $\Delta x = \Delta z = 0.5
c/\omega_{pi}$, time step $\Delta t = 0.01 \Omega_{ci}^{-1}$,
the ratio between light speed and upstream Alfven speed
$c/v_{A1} = 8696.0$, and the anomalous resistivity is taken to be $\eta = 1
\times 10^{-5} 4\pi\omega_{pi}^{-1}$. The initial spatially uniform thermal ion
distribution was generated using $40$ particles per cell. 
Under these parameters, the upstream Alfven speed is about $34.5$ km/s and
the shock speed is about $193$ km/s in upstream frame.


In the second part of our calculation we integrate the motion equation of
an ensemble of test-particle electrons in the electric and magnetic fields
obtained in the hybrid simulations. We assume non-relativistic motions, which is reasonable because
the highest energy of electrons in our study is still non-relativistic.
We release a shell distribution of electrons with energy of $100$ eV, which
corresponds to an electron velocity $V_e = 30.7 U_1 = 5.7 v_{the}$ in the
upstream frame, where $U_1$ is upstream bulk velocity in the shock frame and
$v_{the}$ is the thermal velocity of fluid electrons considered in the hybrid
simulations, respectively. We use a second-order spatial
interpolation and a linear temporal interpolation, which ensure
the smooth variations of the electromagnetic fields.  The test-particle
electrons are released uniformly upstream when the
shock has fully formed and is far from the boundaries. The numerical technique
used to integrate electron trajectories is the so-called Bulirsh-Stoer method. It is highly accurate and
conserves energy well. It is fast when fields are smooth compared with the
electron gyroradius. The algorithm uses an adjustable time-step method based on
the evaluation of the local truncation error. The time step is allowed to vary
between $5 \times 10^{-4} $ and $0.1 \Omega_{ce}^{-1}$, where $\Omega_{ce}$ is
the electron gyrofrequency. The ratio $\Omega_{ce}/\Omega_{ci}$ is taken to be
the realistic value $1836$. The total number of electrons in the simulation is
$1.6 \times 10^6$. The electrons that reach the left or right boundary are
assumed to escape from the shock region and are removed from the simulation. The
boundary condition in the $z$ direction is taken to be periodic. The readers
are referred to \citep{Burgess2006,Guo2010} for more details on the numerical methods.

\begin{figure}
  \includegraphics[width=0.75\textwidth, height=.4\textheight]{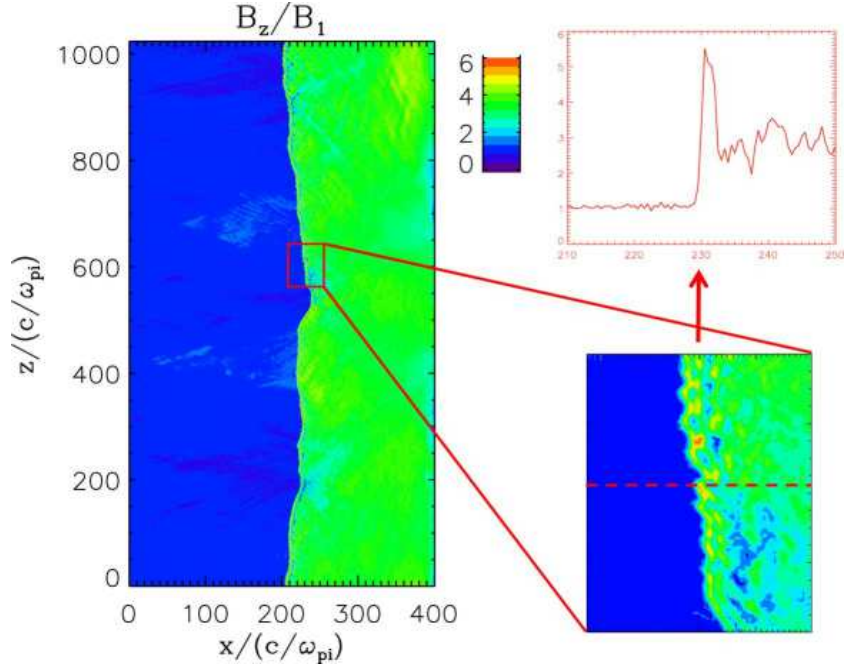}
  \caption{A snapshot of the magnetic field in the z-direction $B_z/B_1$ represented in color-coded scale at $t = 110\Omega_{ci}^{-1}$, where $B_1$ is the averaged upstream magnetic field strength. A region at the shock front is zoomed in on the right bottom and a profile is illustrated on the right upper panel. The shock surface is shown to be rippled and irregular in different scales.}
\end{figure}

\section{Simulation Results}

Figure 1 shows a snapshot of the $z$ component of the magnetic field,
$B_z/B_1$, at $t = 110 \Omega^{-1}_{ci}$.  At this
time, the shock is fully developed. The
position of the shock front is clearly seen from the boundary of the magnetic
field jump. On the right bottom a small 
region of the simulation domain is zoomed in, which shows small scale irregularities at the shock 
front. Because of the effect of the large-scale turbulence, the shock surface become
irregular on a variety of spatial scales from small-scale ripples to large-scale
structures caused by the interaction between the shock and the upstream turbulence
\citep{Giacalone2005a,Lu2009,Guo2010}. Locally, the structure of the shock, shown in top 
right panel, is still clearly a quasi-perpendicular shock. 
The meandering of field lines close to the shock
surface helps to trap the electrons at the shock, leading to efficient
acceleration. The shock ripples also contribute to the acceleration by
mirroring electrons between them. This is shown by examining the trajectories of some electrons 
as illustrated in Figure 2. 
In this figure, the top left plot displays the trajectory of a representative electron in the $x-z$ plane,
overlapped with a 2-D gray-scale representation of $B_z$ at $\Omega_{ci}t =
89.0$. The upper right plot shows
the position of this electron (in $x$) as a function of time. The electron
bounces back and forth between the ripples for several times. The energy change
as a function of the $x$ position, corresponding to these reflections is shown in
the bottom left panel. We find that there are jumps in energy at each of the
reflections. The panel on the bottom right shows the electron energy as a
function of time that also illustrates the features of multiple accelerations
related to the multiple reflections. Since the electrons mainly follow along the 
magnetic lines of force, the large-scale braiding of field lines in space allows 
the fast-moving electrons to interact with the shock front multiple times. 
The electrons can also travel along the meandering magnetic 
field and gain energy from the velocity difference between upstream and downstream \citep{Jokipii2007}.

\begin{figure}
  \includegraphics[width=0.7\textwidth, height=.40\textheight]{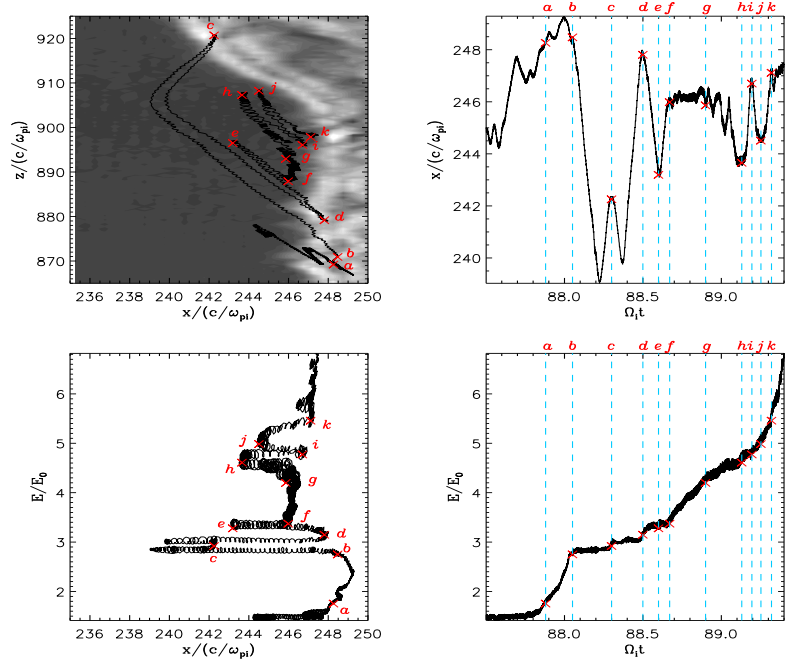}
  \caption{An electron trajectory analysis. The top left panel shows the trajectory of the electron in the $x-z$ plane, overlapped with a contour of $B_z$ magnetic field; The top right panel shows the position of the electron in the $x$ direction as a function of time; The bottom left panel illustrates the electron energy $E/E_0$ as a function of $x$; The bottom right panel shows the dependence of electron energy $E/E_0$ on time.}
\end{figure}

\begin{figure}
\begin{tabular}{ccc}
\epsfig{file=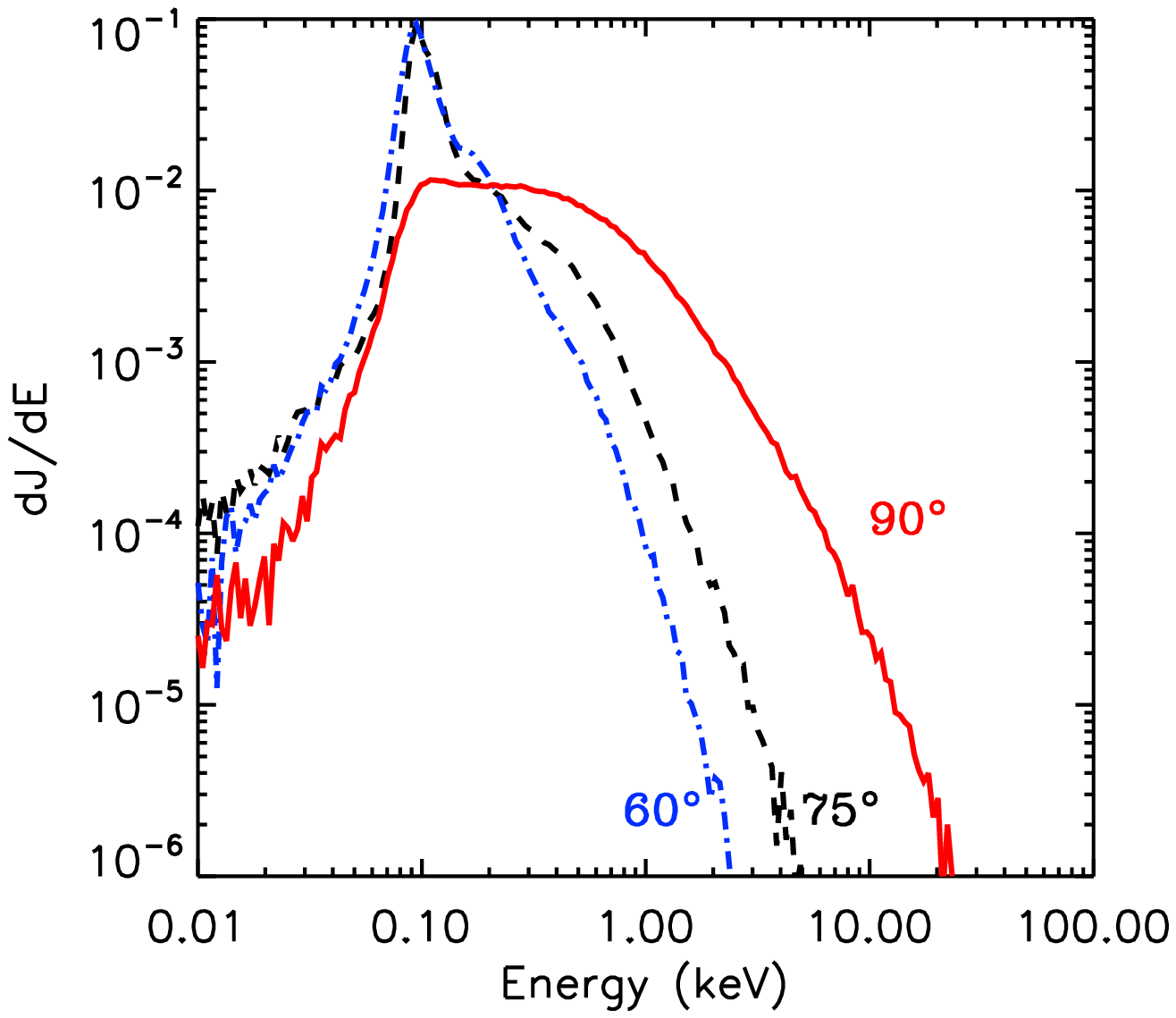,width=12pc} &
\epsfig{file=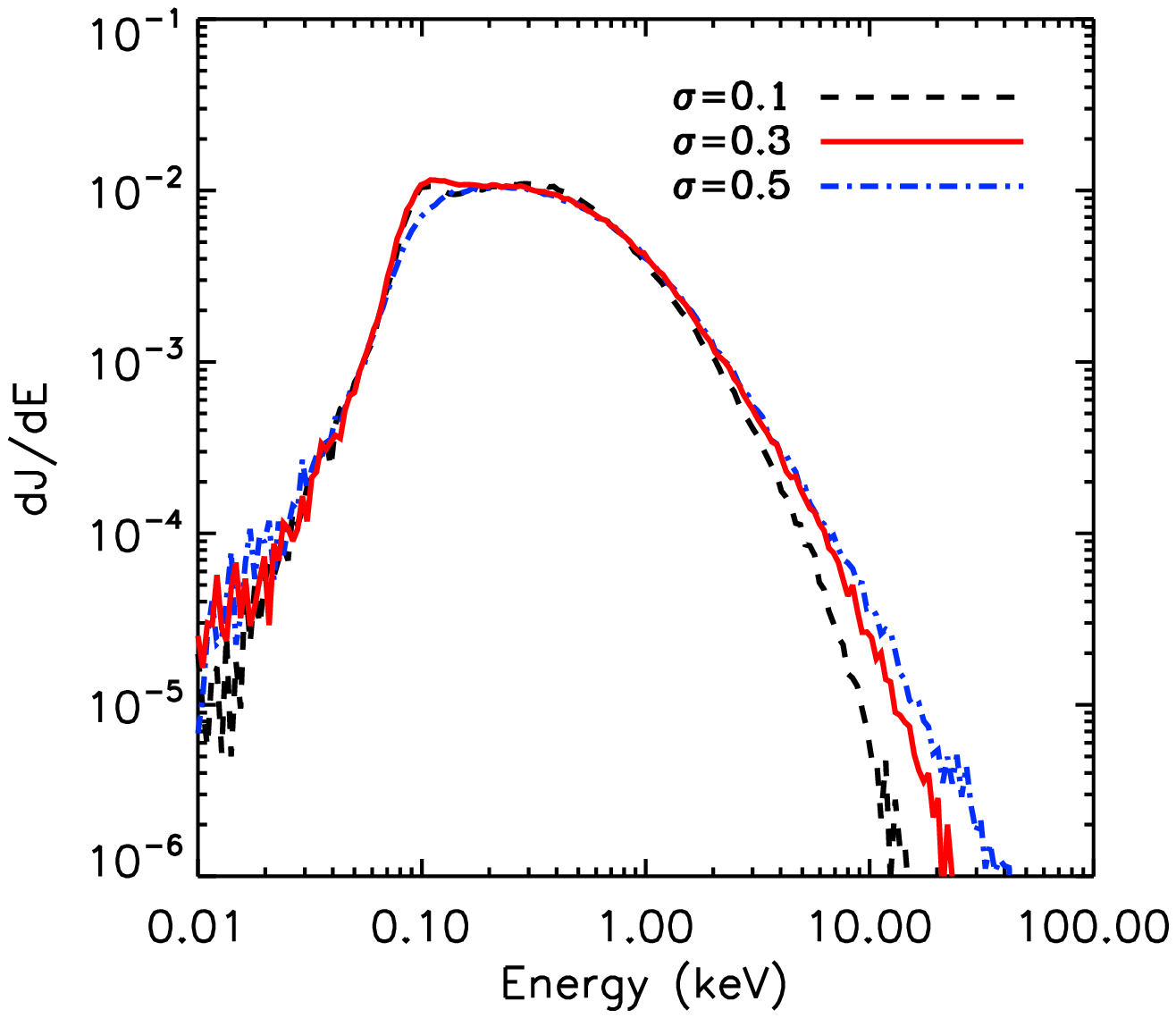,width=12pc} &
\epsfig{file=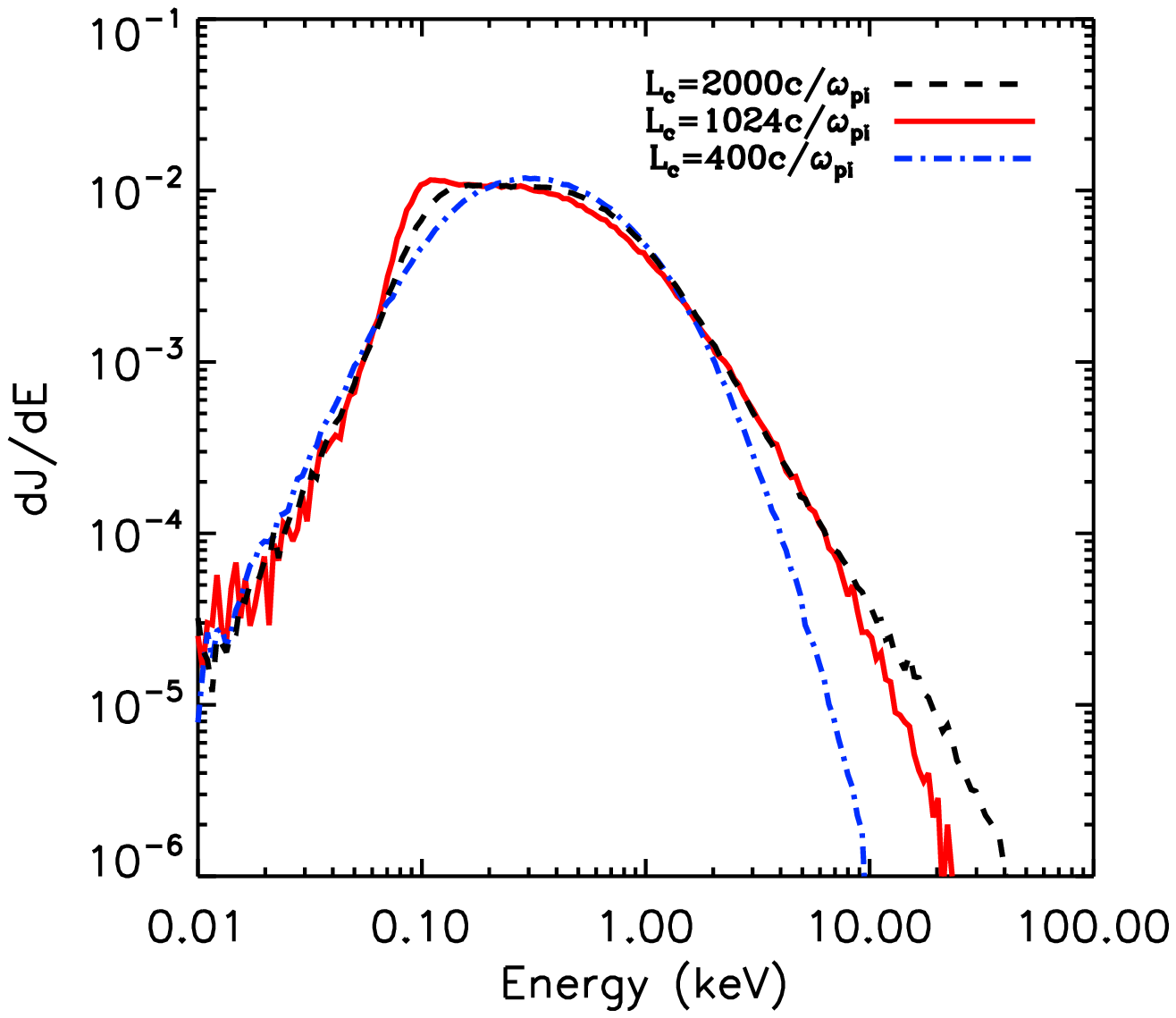,width=12pc} \\
\epsfig{file=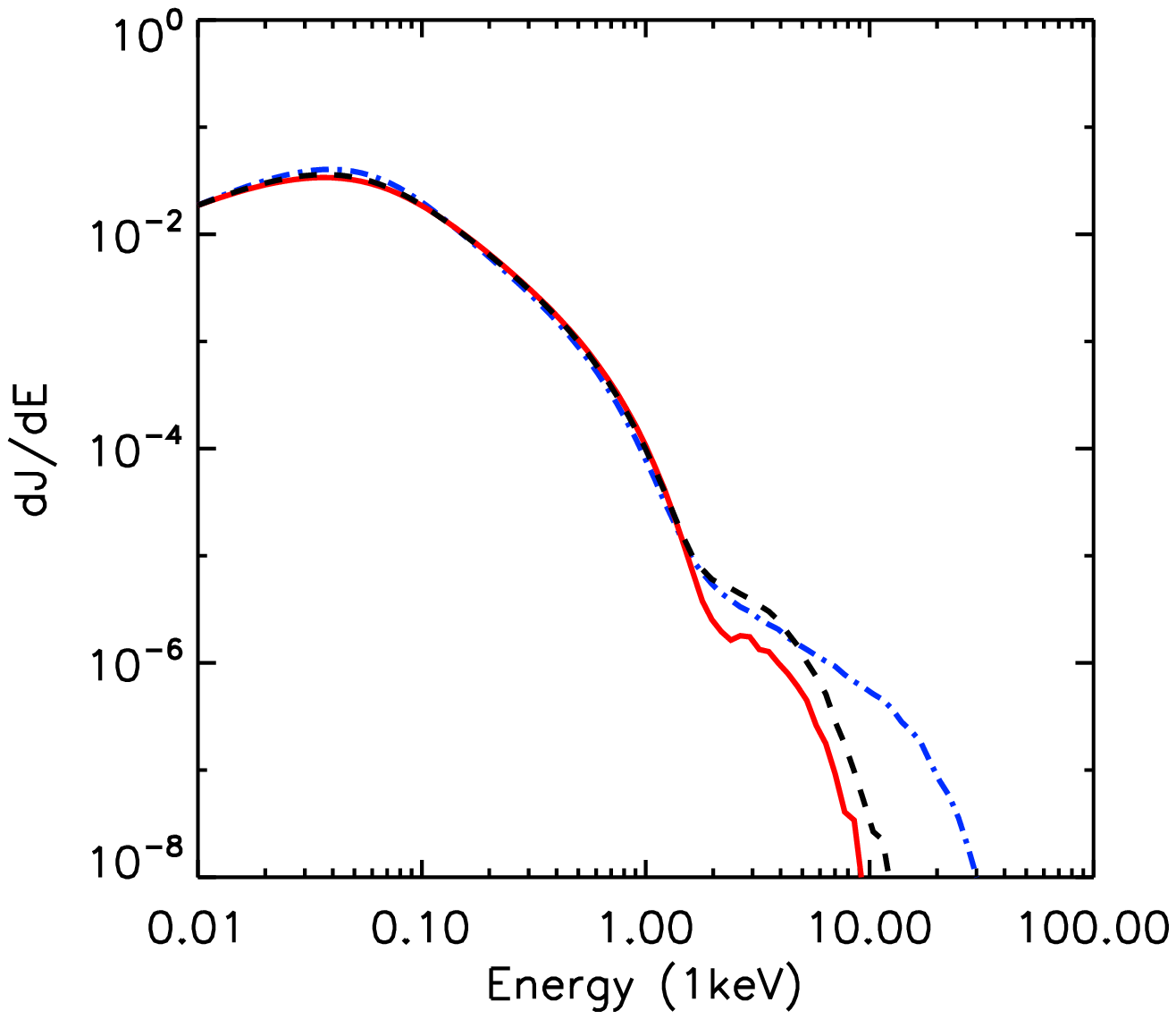,width=12pc} &
\epsfig{file=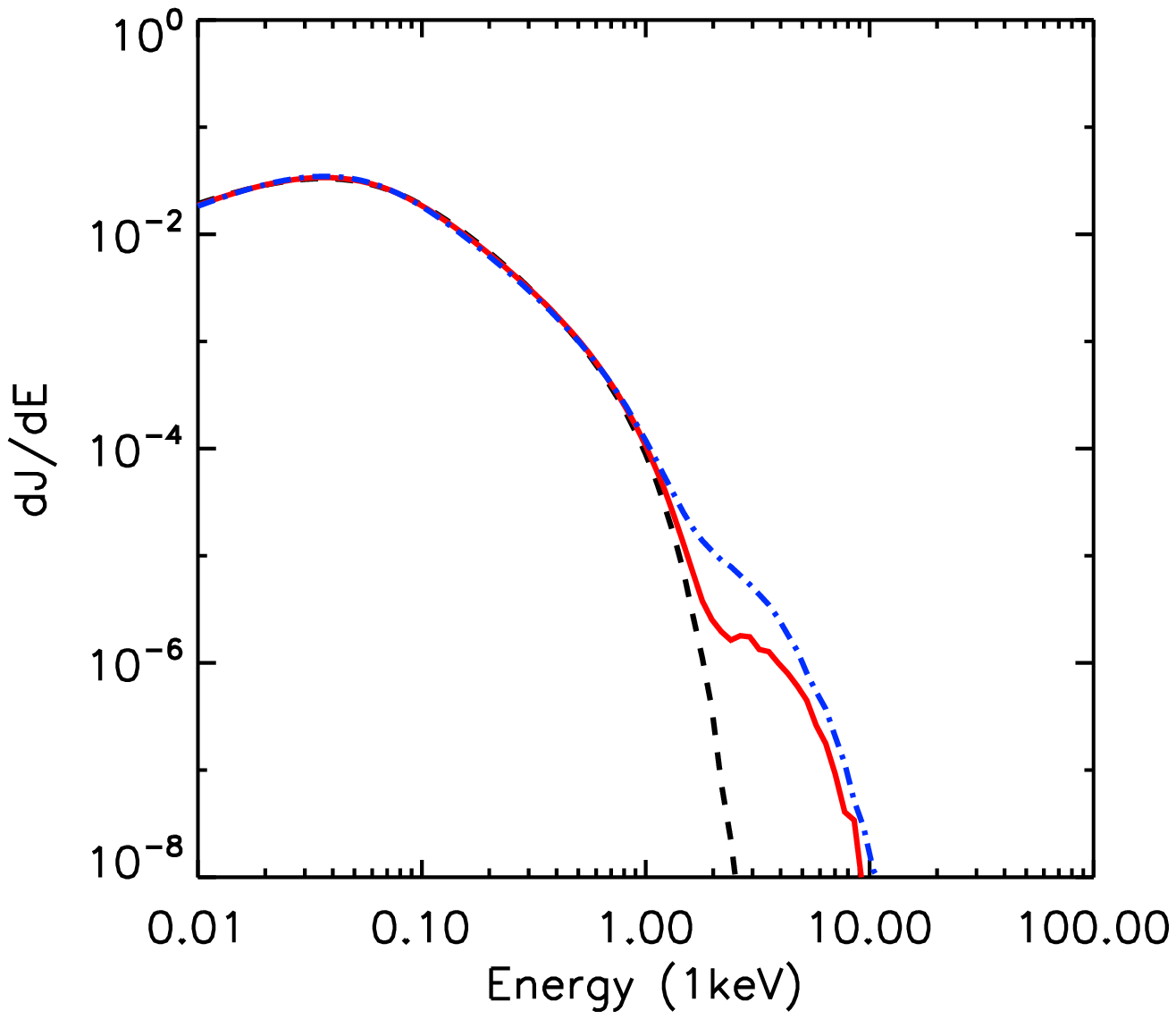,width=12pc} &
\epsfig{file=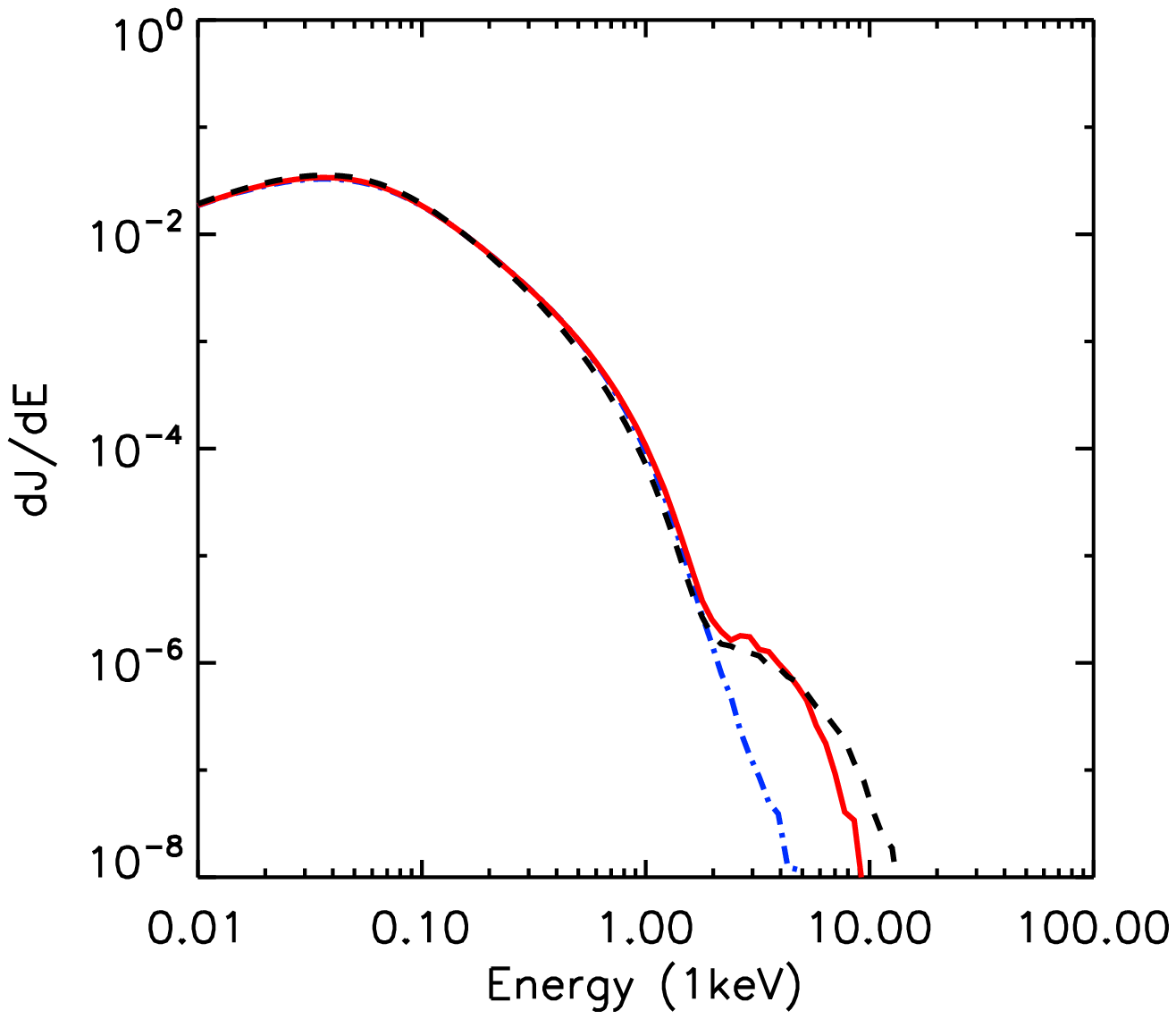,width=12pc} 
\end{tabular}
\caption{Top: energy spectra of accelerated electrons from test-particle electron simulations. Bottom: The energy spectra of protons directly from hybrid simulations. See description in the text for details.}
\label{field}
\end{figure}

We now discuss the effect of varying the averaged shock normal angle, the value of turbulence variances, 
and the values of correlation lengths. In the top panels of Figure 3 we present the electron energy spectra
for left: three different mean shock-normal angles ($\langle \theta_{Bn} \rangle= 60^\circ, 75^\circ$
and $90^\circ$, respectively), middle: three cases with different turbulence variances $\sigma = 0.1$, $0.3$, and $0.5$, respectively, and right: three cases with different sizes of the simulation box and correlation length $L_z = 2000$, $1024$, and $400 c/\omega_{pi}$. All the spectra
are obtained in the end of the simulation $\Omega_{ci}t = 120$ and do not change any more. 
We find this process prefer a perpendicular shock, in which more electrons are accelerated to high energy. It is also found that the energy spectra harden at high energies when the turbulence variance and/or turbulence correlation length is larger, which indicates that the large-scale turbulence is
important for accelerating electrons to high energies. The reason is that the meandering
of field lines is enhanced, which allows the electrons have a better chance to
travel though the shock multiple times.

We explicitly compare the energy spectra of electrons with that of protons. In the bottom panels of Figure 3 we show the energy spectra of downstream protons from the hybrid simulations, corresponding to the spectra of electrons in the top panels. We find more efficient acceleration for protons can be obtained in the case of larger values of turbulence variances and correlation lengths. These agree well with the characteristics of the acceleration of electrons. In our simulations, the accelerated protons at the oblique shock with $\langle \theta_{Bn} \rangle= 60^\circ$ are found to reach higher energies, which is different than previous works \citep{Jokipii1982,Jokipii1987,Giacalone2005a,Giacalone2005b}. This is probably due to the limited temporal and spatial scales of our simulations. As shown by \citet{Giacalone2005b} using test-particle simulations, the energy spectra of protons reach the highest energy in perpendicular shock case in a longer time scale $\Omega_{ci}t \sim 50000$ (this corresponds to $5-10$ minutes for typical parameters in solar corona). However the current results from hybrid simulations do show a population of thermal protons can be accelerated to high energies in perpendicular shocks, which supports the idea that both electrons and protons can be efficiently accelerated by shocks with large shock normal angles.

\section{Summary}

We have presented the results of the acceleration of electrons (and also protons) at a perpendicular shock that propagates through a turbulent magnetic field. The acceleration of electrons are enhanced due to the effect of large-scale magnetic turbulence. The accompanying results for protons qualitatively show the correlation between accelerated electrons and accelerated ions in oblique shocks with large shock normal angles. Since the tight correlation between electrons and ions has been observed in SEP events. This result indicates that quasi-perpendicular/perpendicular shocks play an important role in SEP events.  

\begin{theacknowledgments}
The authors thank Dr. J. R. Jokipii and Dr. Jozsef Kota for valuable discussions and Dr. David Burgess for sharing the details about the simulations method. The authors also thank Dr. Ed Cliver, Dr. Dennis Haggerty and Dr. Allan Tylka for discussions and informations regarding the observations of SEP events related to this project. This work was supported by NASA under grants
NNX10AF24G and NNX11AO64G.
\end{theacknowledgments}

\bibliographystyle{aipproc}   


\begin{thebibliography}{9}

\bibitem[Haggerty
\& Roelof(2009)]{Haggerty2009} Haggerty, D.~K., \& Roelof, E.~C.\ 2009, in AIP Conf. Proc. 1183, Shock waves in Space and Astrophysical Environments, ed. X. Ao, R. H. Burrows, \& G. P. Zank (Melville, NY: AIP), 3

\bibitem[Cliver(2009)]{Cliver2009} Cliver, E.~W.\ 2009, Central 
European Astrophysical Bulletin, 33, 253 

\bibitem[Tylka (2012)]{Tylka2012} Tylka, A. J., private communication in March 2012.

\bibitem[Shih et al.(2009)]{Shih2009} Shih, A.~Y., Lin, R.~P.,
\& Smith, D.~M.\ 2009, The Astrophysical Journal, 698, L152
  
  \bibitem[Jokipii(1982)]{Jokipii1982} Jokipii, J.~R.\ 1982, The Astrophysical Journal, 
255, 716 
  
  \bibitem[Jokipii(1987)]{Jokipii1987} Jokipii, J.~R.\ 1987, The Astrophysical Journal, 
313, 842 

\bibitem[Giacalone(2005)]{Giacalone2005a} Giacalone, J.\ 2005, The Astrophysical Journal,
624, 765

\bibitem[Giacalone(2005)]{Giacalone2005b} Giacalone, J.\ 2005, The Astrophysical Journal,
628, L37

\bibitem[Jokipii 
\& Giacalone(2007)]{Jokipii2007} Jokipii, J.~R., \& Giacalone, J.\ 2007, The Astrophysical Journal, 660, 336 

\bibitem[Guo
\& Giacalone(2010)]{Guo2010} Guo, F., \& Giacalone, J.\ 2010, The Astrophysical Journal, 715, 406

\bibitem[Guo 
\& Giacalone(2012)]{Guo2012} Guo, F., \& Giacalone, J.\ 2012, The Astrophysical Journal, 753, 28 

\bibitem[Burgess(2006)]{Burgess2006} Burgess, D.\ 2006, The Astrophysical Journal, 653, 316 

\bibitem[Lu et al.(2009)]{Lu2009} Lu, Q., Hu, Q., \& Zank, G.~P.\ 2009, The Astrophysical Journal, 706, 687










\end{thebibliography}


\end{document}